# HYPERREACTIVE MODEL IN DYNAMICS OF A VARIABLE-MASS POINT

## Tertychny-Dauri V.Yu.

St.-Petersburg State University of Information Technologies, Mechanics & Optics

Department of Physics and Engineering

tertychny-dauri@mail.ru

49 Kronverkskiy pr., St-Petersburg 197101, Russia

Basing on a new approach to the fundamental conception of the momentum of a variable-mass point, the paper deals with the hyperreactive model of motion. The equations of motion are different in this model from the known Meshchersky-Tsiolkovsky equations, which cannot be taken as a principle in the final analysis by reason of their conflicting behaviour. The suggested conception of hyperreactive motion allows to determine in what way the parameters of motion are dependent on the type of mass change, and thus to solve a basic problem how high absolute velocities may be reached in space.

**1. Introduction.** If one is founded on the proposition that the generalized (Lagrange) coordinates and the mass $M(t)$ of a point are independent at a given instant $t \in [\, t_0,\, t\,]$ for the mass be thus considered as one more Lagrange coordinate (Tertychny-Dauri, 2005, 2008), the total momentum (or measure of motion) of the point can be written in the form

$$Q(t) = M(t)\frac{dR(t)}{dt} + \frac{dM(t)}{dt}R(t), \qquad (1)$$

where $R(t) = r(t) - \rho(t)$ is the peake-to-peake amplitude vector, $dr(t)/dt = v(t)$ is the absolute velocity of the point, $d\rho(t)/dt = u(t)$ is the absolute rate of the radiating center particle outflow, which is a vector function given on $[\, t_0,\, t\,]$. We shall distinguish the quantity $Q(t)$ from a classical momentum $Q_c(t) = M(t)dR(t)/dt$, where the velocity $u(t)$ is assumed, as a rule, stationary $(du(t)/dt = 0)$. From relation (1), one obtains a new universal differential law of dynamics (the principle of completeness, Tertychny-Dauri, 2004, 2006):

$$\frac{dS(t)}{dt} = Q(t),$$

where the vector quantity $S(t) = M(t)\,R(t)$, will be referred to as the motion composition. The principle of completeness or the theorem of changing the motion composition, can be treated as a starting fundamental dynamic law which implies all remainder.

Apply the law of changing the momentum to $Q(t)$. Then the nonstationary equation of point motion will take the form

$$M\frac{dv}{dt} = F + M\frac{du}{dt} - 2\frac{dM}{dt}\frac{dR}{dt} - \frac{d^2M}{dt^2}R, \qquad (2)$$

where $F$ is the applied force, $-(dM/dt)(dR/dt) = (dM/dt)V$ is the standard reactive force (Meshchersky, 1897, Levi-Chivita, 1928), $V = u - v$ is the relative rate of particle outflow, $-R(d^2M/dt^2)$ is the hyperreactive force. Equation (2) is essentially different from the known Meshchersky equation whose derivation is founded on an improper hypothesis of fast responce. To be important in principle, a new model of motion (2) keeps in mind the hyperreactive $d^2M/dt^2$ — containing summand other than the $M$, $dM/dt$ — containing terms. Judging from the new concept of variable-mass point movement, the present work investigates a number of problems which are concerned with the concrete conditions of realizing this movement, brings out characteristic features and corresponding dynamic regularities.

**2. The first problem and Tsiolkovsky's hypothesis**. Let the variable-mass point is moving in a nonresisting medium straightly if no forces are applied and the relative velocity of leaving particles $V$ is assumed constant, collinear to the peake-to-peake amplitude vector $R$, and directed opposite to the point motion. The whole set of these conditions forms, as known, Tsiolkovsky's hypothesis. It is required for the point movement to be determined.

Note at once that the logarithmic relation (Tsiolkovsky, 1914):

$$v = v_0 + \ln(M_0/M), \qquad v_0 = v(t_0), \qquad M_0 = M(t_0)$$

which results from integrating the Meshchersky equation

$$M \frac{dv}{dt} = V \frac{dM}{dt},$$

is not lawful, since $V = const$, $u = const$, and $v = const$.

Make use of scalar equation (2). With this assumption, we obtain

$$\frac{dR}{dt} = -\frac{R\, d^2M/dt^2}{2\, dM/dt}. \tag{3}$$

On integrating (3) with respect to $R$, we have

$$R = R_0 \exp\left\{-\frac{1}{2}\ln\frac{dM/dt}{(dM/dt)_0}\right\},$$

where $R_0 = R(t_0)$, $(dM/dt)_0 = (dM/dt)_{t=t_0}$, whence it appears that

$$\left(\frac{R}{R_0}\right)^2 = \frac{(dM/dt)_0}{dM/dt}. \tag{4}$$

Substitute (4) into (3). Then it takes place

$$\frac{dR}{dt} = -\frac{R_0 (dM/dt)_0^{1/2} (d^2M/dt^2)}{2(dM/dt)^{3/2}}. \tag{5}$$

To be definite, the point mass $M(t)$ will be considered a continuous, monotonously descreasing function of time $\left(dM/dt < 0,\ d^2M/dt^2 > 0,\ [(dM/dt)_0/(dM/dt)]^{1/2} > 0\right)$. Therefore, on the right of relation (5) there is a positive value, and the velocity of the point, $v$, increases obeying the relation

$$v = u - \frac{K_0 d^2M/dt^2}{(dM/dt)^{3/2}}, \qquad (6)$$

where $K_0 = R_0 (dM/dt)_0^{1/2}/2$. The implementation of relation (6) is provided by the mass change following a certain $M = M(t)$ law to be given beforehand; the point velocity $v(t)$ will be determined if $u(t)$ in (6) is a known function of time. If also the relative velocity $V = const$ is given, the mass must change following the differential rule

$$K_0 \frac{d^2M}{dt^2} = V \left(\frac{dM}{dt}\right)^{3/2}$$

This equation may be rewritten in the form

$$\frac{d^2M}{dt^2} - \frac{2V}{R_0 - V(t-t_0)} \frac{dM}{dt} = 0$$

whence, after integrating, we shall obtain the law of mass variation

$$M(t) = \frac{R_0 M_0 + C(t - t_0)}{R_0 - V(t - t_0)} \qquad (7)$$

Where $C = R_0 (dM/dt)_0 - VM_0$ or

$$M(t) = M_0 + \frac{R_0 (dM/dt)_0}{\frac{R_0}{t - t_0} - V}, \qquad \left(\frac{dM}{dt}\right)_0 < 0.$$

**3. The variation of mass.** The time-dependence of mass is dictated by the conditions of engine operation. Make now clear what mass variation ensures the constancy of reactive and hyperreactive forces in the strainght-line motion. Let

$$-\frac{dM(t)}{dt} \frac{dR(t)}{dt} = C_r \qquad (8)$$

be given, where $C_r = const \neq 0$. Since $dR(t)/dt = -V(t)$ then, integrating (8) over $[t - t_0]$ we obtain

$$M(t) = M_0 + C_r \int_{t_0}^{t} \frac{ds}{V(s)}. \qquad (9)$$

Discuss formula (9). We have $V(t) \neq const$ in this relation. Really, if otherwise supposed, a linear mass variation is obtained, whence we have $d^2M(t)/dt^2 = 0$. We have $dV/dt = 0$ on the left of a basal equation of motion (2). If the $F = 0$ condition is fulfilled, we are thus led to an inconsistency and incompetency of Tsiolkovsky's hypothesis.

Consider the case when

$$-\frac{d^2M(t)}{dt^2}R(t) = C_h, \qquad (10)$$

where $C_r = \text{const}$, $R(t) \neq \text{const}$. Twice integrating equation (10), we derive

$$M(t) = M_0 + \left(\frac{dM}{dt}\right)_0 (t - t_0) - C_h \int_{t_0}^{t}\int_{t_0}^{s} \frac{d\omega ds}{R(\omega)}$$

As far as two cases of mass variation are most attractive in good many papers:

1) a linear law $M(t) = M_0[1 - \alpha(t - t_0)]$ and
2) an exponential law $M(t) = M_0 exp[-\alpha(t - t_0)]$, where $\alpha = \text{const} > 0$, we pay attention to these cases too.

If $M(t) = M_0[1 - \alpha(t - t_0)]$, the mass discharge per second is equal to $dM/dt = -\alpha M_0$, where the $\alpha$ parameter is referred to as specific mass discharge per second; in addition, we hav$d^2M(t)/dt^2 = 0$.e

Let $F = 0$ be given. Then the Tsiolkovsky hypothesis on a constant $V$ does not fulfilled, and the relative velocity must comply with the equation

$$M\frac{dV}{dt} + 2V\frac{dM}{dt} = 0,$$

i. e.

$$V(t) = \frac{V_0}{[1 - \alpha(t - t_0)]^2}, \quad R(t) = R_0 - \frac{V_0(t - t_0)}{1 - \alpha(t - t_0)}, \qquad (11)$$

Where $V_0 = V(t_0)$. Let $a_r$ stand for the acceleration due to the action of doubled reactive force. Then, if the law of mass variation is linear and with (11), we have

$$a_r = \frac{2V\, dM/dt}{M} = \frac{2\alpha V_0}{[1 - \alpha(t - t_0)]^3},$$

And $n = a_r/g$ is the overload produced by the reactive force, where $g$ is the gravity acceleration.
Consider the exponential law of mass variation $M(t) = M_0 exp[-\alpha(t - t_0)]$. We derive

$$\frac{dM}{dt} = -\alpha M_0 e^{-\alpha(t-t_0)} = -\alpha M, \qquad \frac{d^2M}{dt^2} = \alpha^2 M.$$

Let $F = 0$ be given. The the Tsiolkovsky hypothesis is readily seen not to be fulfilled for the exponential law
(the equations are inconsistent). We now have

$$M\frac{d^2R}{dt^2} = -2\frac{dM}{dt}\frac{dR}{dt} - \frac{d^2M}{dt^2}R,$$

or

$$\frac{d^2R}{dt^2} - 2\alpha\frac{dR}{dt} + \alpha^2 R = 0, \qquad (12)$$

upon integrating equation (12), we obtain

$$V(t) = [V_0 + \alpha(V_0 + \alpha R_0)(t - t_0)]e^{\alpha(t-t_0)}, \quad R(t) = [R + (\alpha R_0 + V_0)(t - t_0)]e^{\alpha(t-t_0)},$$

whence we find the following expressions for the corresponding acceleration and overload:

$$a_r = -2\alpha V, \quad a_h = -\alpha^2 R, \quad a_{r+h} = -\alpha(2V + \alpha R), \quad n = a_{r+h}/g.$$

**4. The second Tsiolkovsky problem.** Analyse the motion of variable-mass point in a homogeneous gravity field when the point moves vertically upwards. It is required to find the law of mass and distance variation as a function of time and the maximal point altitude. Being directed vertically downwards, the relative velocity of the emitted particles is constant.
In this case, the equation of motion has the form

$$\frac{dR}{dt} = -\frac{R d^2 M/dt^2}{2 dM/dt} - \frac{Mg}{2 dM/dt}. \tag{13}$$

We integrate (13) and obtain

$$R(t) = \frac{(dM/dt)_0^{1/2}}{(dM/dt)^{1/2}}\left[R_0 - \frac{g}{2(dM/dt)_0^{1/2}} \int_{t_0}^{t} \frac{M\,dt}{(dM/dt)^{1/2}}\right]. \tag{14}$$

If (14) is substituted into (13), we shall have the expression for $dR/dt$ as a function of $M(t)$, $dM/dt$, and $d^2 M(t)/dt^2$, whence we derive

$$v = u - \frac{(dM/dt)_0^{1/2}(d^2M/d^2t)}{2(dM/dt)^{3/2}}\left[R_0 - \frac{g}{2(dM/dt)_0^{1/2}} \int_{t_0}^{t} \frac{M\,dt}{(dM/dt)^{1/2}}\right] - \frac{Mg}{2 dM/dt}, \tag{15}$$

Where $u(t)$, $M(t)$ are the given functions of time. If given is a constant relative velocity $V$, then equation (13) provides a differential law of mass variation in the form

$$\frac{d^2 M}{dt^2} - \frac{2V}{R_0 - V(t-t_0)}\frac{dM}{dt} - \frac{gM}{R_0 - V(t-t_0)} = 0. \tag{16}$$

Equation (16) is not integrable in quadratures. It is to be noted that (16) is the relative-equilibrium equation for a variable-mass point in a rgavity field if one assumes $V = u = const$.
Determine for the point its maximal altitude $H$. As far as we have $v(t_*) = 0$ for this altitude value, the time of rising, $t_*$, may be found from equation (15), and the following relation,

$$H = \rho_0 + \int_{t_0}^{t_*} u(s)ds + R(t_*)$$

is valid, where $u(t)$, $M(t)$ are the given functions of time, $R(t_*)$ being determined by formula (14) taken for $t = t_*$.

Compare the result obtained to the case when $M(t) = M_0 \exp[-\alpha(t-t_0)]$ and $V \neq const$. We have

$$M\frac{d^2 R}{dt^2} = -2\frac{dM}{dt}\frac{dR}{dt} - \frac{d^2 M}{dt^2}R - Mg$$

or

$$\frac{d^2 R}{dt^2} = -2\alpha\frac{dR}{dt} + \alpha^2 R = -g,$$

whence there takes place

$$R(t) = [R_0 - (V_0 + \alpha R_0)(t - t_0)]e^{\alpha(t-t_0)} - \frac{g}{\alpha^2}, \quad V(t) = [V_0 + \alpha(V_0 + \alpha R_0)(t - t_0)]e^{\alpha(t-t_0)}, \quad (17)$$

and the time $t_*$ is found from the equation

$$u(t_*) = [V_0 + \alpha(V_0 + \alpha R_0)(t_* - t_0)]e^{\alpha(t_* - t_0)}$$

and then we led to

$$H = \rho_0 + \int_{t_0}^{t_*} u(s)ds + R(t_*).$$

The case when the mass decreases following the linear law, may be considered analogously.

**5. Optimal conditions of motion in the Tsiolkovsky problems.** It was shown in the foregoing sections that basic characteristics of the point motion depend on the law of its mass variation. Therefore, it is worth saying of the formation of optimal conditions of motion. Denote the fuel reserve, $M(t) = N + m(t)$, where $N = $ const is the mass of the object excluding fuel (body, engine, equipment), by $m(t)$. It follows from relations (6) and (15) that the velocity acquired by the point depends on the rate $(dM/dt = dm/dt$ and the acceleration $(d^2M(t)/dt^2 = d^2m(t)/dt^2)$ in he fuel-mass variation. This is a very important conclusion. Relation (6) does not contain the mass at all. Try to foresee this result from mere qualitative considerations. According to the previous theory, the more fuel mass of the object, the more is its velocity. In this case, it does not matter of what nature the quantities $dm/dt$, $d^2m(t)/dt^2$ are. The contrary is really the case: to allow for the fuel mass with ignoring these quantities means, in principle, to allow, as fast as the object moves, for the impact effects only (it is borne in mind a disconcentrated and distributed impact effect). But in fact, as shown by a new theory, the small fuel mass can provide for marked and necessary bsolute velocities. Let us return to the first Tsiolkovsky problem. When the fuel mass $m(t)$ becomes zero at $t = t_*$, the length of active section will be

$$r(t_*) = \rho_0 + \int_{t_0}^{t_*} u(s)ds + \frac{R_0(dM/dt)_0^{1/2}}{(dm/dt)_*^{1/2}}, \quad (18)$$

Where $t_*$ is determined from the equation $M(t_*) = N$. If the relative rate $V = $ const is given, then $t_*$ and $(dM/dt)_*$ are estimated with the use of formula (7):

$$t_* = t_0 + \frac{m_0 R_0}{m_0 V - R_0 (dm/dt)_0}, \quad \left(\frac{dM}{dt}\right)_* = \left(\frac{dm}{dt}\right)_* = \frac{R_0^2 (dm/dt)_0}{[R_0 - V(t_* - t_0)]^2}.$$

The case of an instantaneous fuel rejection (a concentrated single impact) corresponds to $t_* = t_0$, i. e. $r(t_*) = r(t_0) = \rho_0 + R_0$,

$$\frac{V}{R} - \frac{(dm/dt)_0}{m_0} \to \infty.$$

But if we have

$$V \to \frac{R_0(dm/dt)_0}{m_0},$$

Then $t_* \to \infty$ : one one observes an inifestimal fuel-mass variation, and in addition, $r(t_*) \to \infty$. Pass to the second Tsiolkovsky problem. We are to find a maximal altitude of active section in rising a variable-mass point which moves in the homogeneous gravitational field vertically upwards in case when the law of changing $M(t)$ is given in advance. We have (for $V = $ const and not given a priori)

$$H = \rho_0 + \int_{t_0}^{t_*} u(s)ds + R(t_*), \quad M(t_*) = N,$$

$$R(t_*) = \frac{(dm/dt)_0^{1/2}}{(dm/dt)_*^{1/2}} \left[ R_0 - \frac{g}{2(dm/dt)_0^{1/2}} \int_{t_0}^{t_*} \frac{M\,dt}{(dm/dt)^{1/2}} \right].$$

This problem will have an analytical solution if the $M(t)$ function of time is known. Let $M(t) = M_0 exp[-\alpha(t-t_0)]$, $M_0 = N + m_0$, $V \neq$ const be given. Resolving the equation $M(t_*) = N$, one finds the time $t_*$ at which the fuel reserve is equal to zero (the end of active section)

$$(N = m_0)e^{-\alpha(t_*-t_0)} = N,$$

whence it follows that

$$t_* = t_0 + \ln\left(1 + \frac{m_0}{N}\right)^{1/\alpha}. \tag{19}$$

Consider equation (17) for $R(t)$: Substituting (19) into (17), one obtains

$$R(t_*) = \left[R_0 - (V_0 + \alpha R_0)\ln\left(1 + \frac{m_0}{N}\right)^{1/\alpha}\right]\left(1 + \frac{m_0}{N}\right) - \frac{g}{\alpha^2}. \tag{20}$$

If (19) and (20) are now substituted into the expression for $H = \rho_0 + \int_{t_0}^{t_*} u(s)ds + R(t_*)$, one will obtain the value of altitude of active section in decreasing the fuel reserve by the exponential law. Determine at what value of $\alpha$ the altitude of point rise will be maximal. Differentiating $H$ with respect to $\alpha$, the equation $\partial H/\partial \alpha = 0$ written to determine the optimal value of $\alpha$, will have the form (after the reduction by $1/\alpha^2$)

$$\ln\left(1 + \frac{m_0}{N}\right) u\left[t_0 + \ln\left(1 + \frac{m_0}{N}\right)^{1/\alpha}\right] = \frac{2g}{\alpha} + V_0\left(1 + \frac{m_0}{N}\right). \tag{21}$$

In particular, the maximum altitude of active section can be achieved if one assumes $\alpha = \infty$ in equation (21), which corresponds to the instantaneous fuel rejection. Whence we have a choice restriction for the initial data: the condition $u(t_0) = V_0(1 + m_0/N)$ must be fulfilled. Emphasize the fact that the previous theory only determines the maximum altitude for $\alpha = \infty$, a new hyperreactive model generates the whole set of solutions to this problem.

The case of a linear decrease in the fuel mass can be considered in the same way, and will not decide upon that. If the value $V =$const is given in advance, then equation (16) is not integrated in quadratures with respect to $M(t)$. However, its numerical solution approximately gives the time $t_*$ for which we have $m(t_*) = 0$, and the maximum altitude of active section will be thus found.

    **6. Conclusions.** Let us attempt to sum up the said above and to formulate the conclusions which are based on the presented here results on the hyperreactive simulation.
1. The conceptual basis of the Meshchersky - Tsiolkovsky model does not held good. First this model is inconsistent at the differential level; second it does not permit to have regard to the $d^2M(t)/dt^2 -$ containing terms, and finally this model does not blend with the Lagrange scheme dynamically describing the object in the generalized undependent coordinates.
2. The hyperreactive model allows to reach the necessary absolute rates mainly at the cost of the rate behaviour and change deceleration of the object mass.
3. The previous theory makes special reference to setting the relative rate of particle outflow, $V$, and the new one gives preference to setting the absolute rate of particle outflow, $u$.
4. As follows from the results, advancing the Tsiolkovsky hypothesis on the fact that $V$ is constant, is unwarranted: the dynamics of point movement is bound.
5. For the hyperreactive model to function effectively and the necessary final velocity to be reached, it is requested for the accelerated particle motion to be provided inside the object, in other words, the acceleration motion of fuel particles inside the engine set. The hyperreactive engine must perform an accelerated motion of fuel particles. The basis of such accelerated engine can be the most diverse physical and technical ideas.
6. At the modern stage, the real-system hyperreactive effect mainly manifests itself in the occurence of transitivity (for example in the accelerated motion) and in the orbit deviation from the calculated one.
7. The new theory can be only borne out by the practical activities. The present paper is of essentially logical and theoretical, preferentially research, nature.